\documentclass{article}
\usepackage{spconf,amsmath,graphicx,array}
\usepackage{booktabs}
\usepackage{multirow}
\usepackage{cite}
\usepackage{url}
\usepackage{hyperref}
\usepackage{tablefootnote}
\usepackage{footnote}
\usepackage[table,xcdraw]{xcolor}

\title{Speech Foundation Model Ensembles for the Controlled Singing Voice Deepfake Detection (CtrSVDD) Challenge 2024}
\name{Anmol Guragain\thanks{\textsuperscript{*}These authors contributed equally to this work. 
}
$^{1,2}$\textsuperscript{*}, Tianchi Liu$^{1}$\textsuperscript{*}, Zihan Pan$^{1}$, Hardik B. Sailor$^{1}$, Qiongqiong Wang$^{1}$}
\address{\fontsize{11pt}{10pt}\selectfont $^{1}$ Institute for Infocomm Research (I$^2$R), Agency for Science, Technology and Research (A$^\star$STAR), Singapore\\
\fontsize{11pt}{10pt}\selectfont $^{2}$ Vellore Institute of Technology, India}

\begin{document}
%
\maketitle
\begin{abstract}
This work details our approach to achieving a leading system with a 1.79\% pooled equal error rate (EER) on the evaluation set of the Controlled Singing Voice Deepfake Detection (CtrSVDD). The rapid advancement of generative AI models presents significant challenges for detecting AI-generated deepfake singing voices, attracting increased research attention. The Singing Voice Deepfake Detection (SVDD) Challenge 2024 aims to address this complex task. In this work, we explore the ensemble methods, utilizing speech foundation models to develop robust singing voice anti-spoofing systems. We also introduce a novel Squeeze-and-Excitation Aggregation (SEA) method, which efficiently and effectively integrates representation features from the speech foundation models, surpassing the performance of our other individual systems. Evaluation results confirm the efficacy of our approach in detecting deepfake singing voices. 
{The codes can be accessed at \url{https://github.com/Anmol2059/SVDD2024}.}


\end{abstract}
\begin{keywords}
Singing voice, deepfake detection, anti-spoofing, SVDD, SSL, SEA
\end{keywords}
\section{Introduction}
\label{sec:intro}
With the rapid development of generative AI technology, the quality of audio synthesis has significantly improved, making it increasingly difficult to distinguish between bona fide and spoofed audio. However, this progress also poses significant risks to human voice biometrics and can deceive both automatic speaker verification systems and their users~\cite{7858696}. Additionally, the proliferation of spoofed speech presents a serious threat to cybersecurity, as it can be used to manipulate information, conduct fraud, and bypass security measures that rely on voice authentication. Finding effective ways to detect spoofing attacks and protect users from the threat of spoofed speech is becoming increasingly important. Therefore, speech anti-spoofing, also known as speech deepfake detection, has emerged~\cite{ASVspoof2019, delgado2024asvspoof, one_class, AASIST, wu2024codecfake}. It is dedicated to developing reliable automatic spoofing countermeasures (CMs), which is of utmost importance to society and the ethical applications of generative models.

Unlike speech spoofing, creating deepfakes of singing voices introduces distinct challenges. This complexity arises from the inherently musical aspects of singing, such as varying pitch, tempo, and emotion, as well as the frequent presence of loud and intricate background music~\cite{SingFake, CtrSVDD_interspeech}. These factors make it more difficult to detect deepfakes in singing compared to regular speech, which typically features a more consistent and predictable sound pattern.
Recently, the speech anti-spoofing research community has been increasingly focusing on this challenging issue, resulting in the development of related datasets~\cite{SingFake, CtrSVDD_interspeech, 10446271}, challenges~\cite{SVDD_challenge}, and models~\cite{chen2024singing}.
The Singing Voice Deepfake Detection (SVDD) Challenge 2024 aims to address these challenges by fostering the development of robust detection systems~\cite{SVDD_challenge, zhang2024svdd2024inauguralsinging}. 

Speech foundation models are large, pre-trained models designed to serve as the backbone for various speech-related tasks, including speaker verification, speech recognition, and more~\cite{lin2024sawavlms, 10446072, jiang2024target}. Many of these models rely on self-supervised learning (SSL) to develop robust speech representations, such as WavLM~\cite{wavlm} and wav2vec2~\cite{wav2vec2}. These models excel in learning high-quality representations that can be fine-tuned for specific downstream tasks. Recently, many studies on speech anti-spoofing have adopted this approach and achieved state-of-the-art performance~\cite{10003971, liu2024neural, 10446331, 9747768, 10448049, 10448016}. The progress of these studies and their promising performance motivate us to continue exploring along this particular line.

This work details our participation in the CtrSVDD track of the SVDD Challenge 2024. 
We detect singing voice deepfakes by ensembling models developed using speech foundation models, data augmentation techniques, and various layer aggregation methods. Specifically,  {the default Weighted Sum aggregation method fixes weights after training, limiting adaptability to new data. The recently proposed Attentive Merging (AttM) method~\cite{attentive_merge}, while powerful, can lead to overfitting on small datasets. To address these issues,} inspired by Squeeze-and-Excitation Networks (SENet)~\cite{SE}, we propose the SE Aggregation (SEA) method. {This method dynamically assigns weights and mitigates overfitting issues, enabling our best individual model to achieve an EER of 2.70\% on the CtrSVDD evaluation set. Further investigations show that ensembling systems enhances robustness and performance, achieving our best result of 1.79\% EER.}


\section{Methodology}
\label{sec:methodology}
\subsection{Data Augmentation}
\label{subsubsec:rawboost_aug}

We employ the RawBoost augmentation~\cite{rawboost}, which introduces various types of noise to the audio data to simulate real-world acoustic variations. These augmentation types include:

\begin{itemize}
\item (1) Linear and non-linear convolutive noise (LnLconvolutive noise). This involves applying a convolutive distortion to the feature set by filtering the input signal with notch filter coefficients, iterating $N_f$ times,, and raising the signal to higher powers to simulate real-world distortions.

\item (2) Impulsive signal-dependent noise (ISD additive noise). This is introduced by adding noise to a random percentage of the signal points, scaled by the original signal's amplitude.

\item (3) Stationary signal independent noise (SSI additive noise). This represents stationary signal-independent noise, which is added uniformly across the signal.


\end{itemize}

\begin{figure*}[t]
\vspace{0.1 in}
\centerline{\includegraphics[scale=0.14]{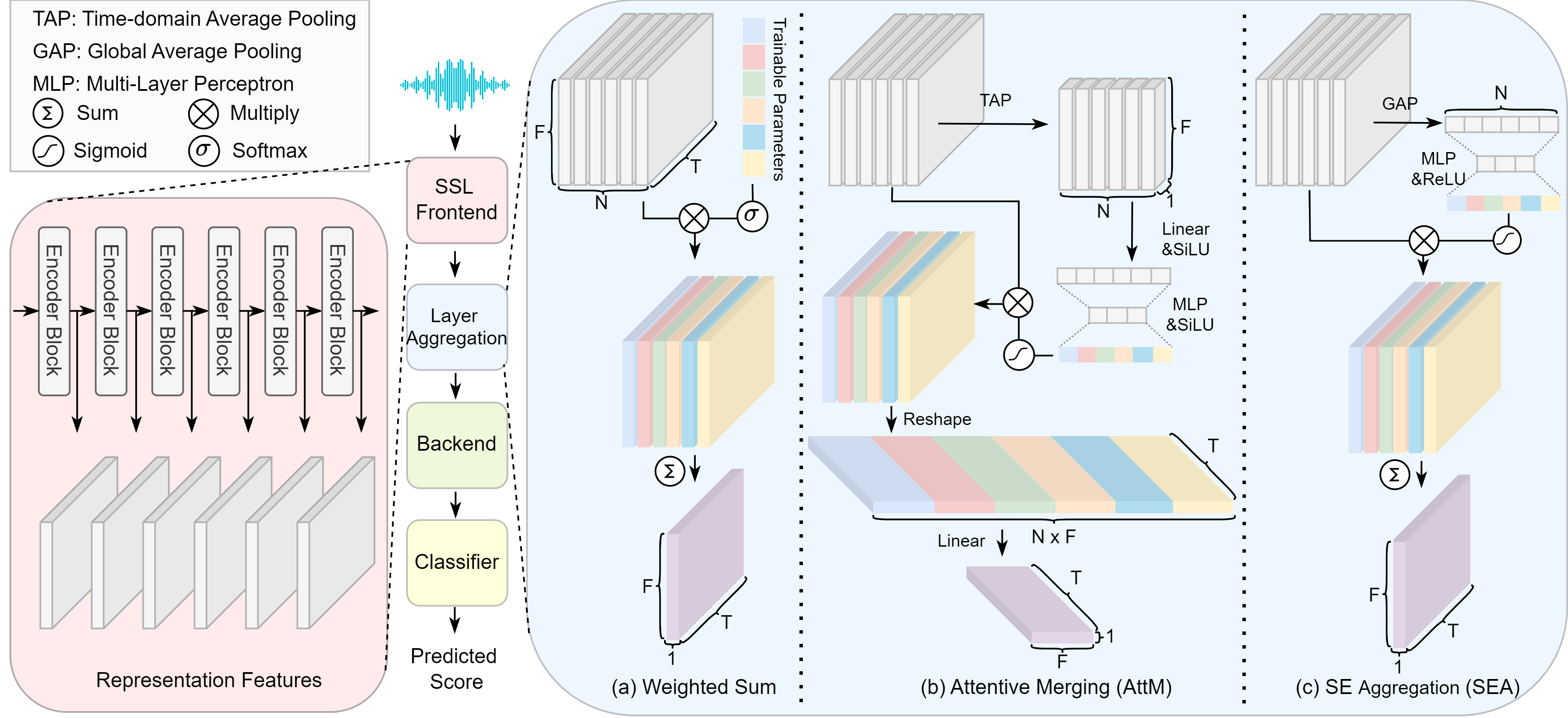}}
\caption{The system architecture of a speech foundation model-based singing voice deepfake detection system. The top-left corner shows the legend. The bottom-left section illustrates the SSL-based front-end, with its output being representation features of  $N \times F \times T$, where 
$N$ is the number of layers in the SSL encoder, 
$F$ is the dimension of the representation features, and $T$ is the number of frames. In this figure, 
$N$ = 6 is used as an example. The right side details the layer aggregation process, including the three aggregation strategies used in this work: (a) Weighted Sum, (b) Attentive Merging (AttM)~\cite{attentive_merge}, and (c) the proposed SE Aggregation (SEA).}
\label{fig_system}
\vspace{0.1 in}
\end{figure*}

\subsubsection{Parallel Noise Addition}

We adopt a parallel noise addition strategy to independently incorporate multiple noise characteristics. We process the input feature through both LnL Convolutive Noise and ISD Additive Noise algorithms simultaneously, resulting in two separate noisy signals. These signals are then combined by summing and normalizing to maintain consistent amplitude levels. This parallel approach allows each noise type to influence the signal independently, effectively capturing the combined effects of convolutive and impulsive noise, and providing a robust simulation of complex noise conditions. This method is referred to as the `parallel: (1)+(2)' approach described in RawBoost~\cite{rawboost}.

\subsubsection{Sequential Noise Addition}
We use a sequential noise addition process to enhance the robustness of our features, incorporating the aforementioned three types of noise. This sequential approach ensures comprehensive noise simulation and results in various combinations such as `series: (1)+(2)', `series: (1)+(3)', and `series: (2)+(3)', following those in RawBoost~\cite{rawboost}.


\subsection{Individual Models Description}
\subsubsection{Frontend}
In this subsection, we provide a detailed overview of the frontends used in our individual models, emphasizing their ability to efficiently process raw audio data.

\textbf{Raw waveform}. Following the baseline system described in the SVDD challenge 2024~\cite{SVDD_challenge}, we employ RawNet2~\cite{jung20c_interspeech}-style learnable SincConv layers with 70 filters as the frontend. These SincConv layers are designed to effectively capture essential features from raw audio signals, enhancing the model's ability to process and analyze audio data for subsequent tasks.

\textbf{wav2vec2}.
The wav2vec2 model offers significant advantages in effectively capturing a wide range of audio features directly from raw audio inputs~\cite{wav2vec2}. This model excels in extracting detailed and nuanced information from audio data, which can then be utilized for various downstream tasks such as speaker verification, speech recognition, and speech anti-spoofing. By processing the raw audio waveforms without requiring extensive pre-processing, wav2vec2 enhances the ability to perform complex audio-related tasks with improved accuracy and efficiency. This direct approach not only simplifies the workflow but also improves the overall performance of the subsequent processing and classification tasks~\cite{truong2024temporal}.

\textbf{WavLM}.
The WavLM~\cite{wavlm} is a large-scale pre-trained speech foundation model for addressing the multifaceted nature of speech signals, including speaker identity, paralinguistics, and spoken content. Its robust performance on the SUPERB benchmark~\cite{yang21c_interspeech} underscores its potential versatility across diverse speech processing applications. Given its advanced capabilities in modeling and understanding complex speech patterns, WavLM holds promise for use in specialized area of singing voice deepfake detection. The model's ability to capture intricate vocal nuances and sequence ordering could be instrumental in identifying synthetic patterns in singing voices, thereby contributing to the SVDD task.

\subsubsection{Layer Aggregation Strategy}
The layer aggregation strategy in speech foundation models refers to the technique of combining information from multiple layers to enhance the model's performance in speech-related downstream tasks like speaker verification, emotion recognition, and anti-spoofing. Each layer in a speech foundation model captures distinct aspects and features of the input waveform. By aggregating these layers, the model can leverage a richer set of features, combining low-level acoustic information from early layers with higher-level semantic and contextual information from later layers. This process typically involves techniques such as concatenation, weighted sum, or attention mechanisms to effectively aggregate the multi-layer representations~\cite{huo23b_interspeech}. These learned weights allow the model to emphasize more relevant features and reduce noise or less important information. In this work, we explore weighted sum and attentive merging (AttM)~\cite{attentive_merge}. Inspired by SE~\cite{SE}, we propose SE Aggregation. These three methods are illustrated in Fig.~\ref{fig_system}, and the details are as follows:

\textbf{Weighted Sum}. The weighted sum method combines outputs from multiple neural network layers using adjustable parameters. Each layer's output receives a unique weight, enabling the model to determine the optimal contribution of each layer to the final representation. These weights are adjusted during the training process to enhance the model's performance and remain fixed during inference.

\textbf{Attentive Merging (AttM)}. The AttM~\cite{attentive_merge} approach emphasizes the most relevant features for anti-spoofing by averaging the embeddings across the time dimension and applying a fully connected layer to squeeze the hidden dimensions. Attentive weights are computed using a sigmoid activation function, which are then applied to the stack of embeddings. Finally, a linear projection network merges these re-weighted embeddings, retaining global spatial-temporal information while emphasizing the most relevant transformer layers for anti-spoofing. This method not only achieves state-of-the-art performance but also improves computational efficiency by utilizing only a subset of the transformer layers~\cite{attentive_merge}.

\textbf{Proposed SE Aggregation (SEA)}. The weighted sum method is simple yet effective. However, its weights are fixed after training, limiting its adaptability to new data. The AttM method, though powerful, requires a large number of parameters, which can lead to overfitting on small datasets. 
Most of these parameters are concentrated in the final linear layer.
To address this, we introduce a new method called SE Aggregation (SEA), inspired by SENet~\cite{SE}, which eliminates the need for the final linear layer. SEA enables a lightweight, cross-layer attention-based aggregation.

The SE module is well-knwon for its ability to adaptively recalibrate channel-wise feature responses by explicitly modeling interdependencies between channels~\cite{SE}. This recalibration enhances the representational capability of the network by focusing on the most informative features and suppressing less useful ones, which is crucial for tasks requiring high precision and robustness~\cite{MFA_TDNN}. This method has been widely applied and validated in speech tasks, such as anti-spoofing~\cite{lai19b_interspeech, 9003763} and speaker verification~\cite{Desplanques20, RecXi, liu2023golden}. Instead of using this approach to re-weight channels, we employ it to compute layer attention, dynamically emphasizing important channels for each sample. The proposed SEA method operates by initially compressing temporal and channel information through a global average pooling (GAP) operation, creating a layer-wise descriptor. This descriptor is then used to selectively emphasize informative features, as illustrated in Fig.~\ref{fig_system} (c).

Notably, the layer aggregation technique is only applied to the speech foundation model-based systems in this work. The RawNet2-based system does not require the layer aggregation strategy.

\subsubsection{Backend}

{The audio anti-spoofing using integrated spectro-temporal graph attention networks (AASIST)} functions as the  model, leveraging graph-based attention mechanisms to capture spectral and temporal audio features~\cite{AASIST}. It includes several key components~\cite{AASIST}:
\begin{itemize}
    \item The {Graph Attention Layer (GAT)} computes attention maps between nodes and projects them using attention mechanisms. This layer consists of linear layers, batch normalization, dropout, and Scaled Exponential Linear Unit (SELU) activation. Separate GAT layers are used for spectral and temporal  features.
    \item The {Heterogeneous Graph Attention Layer (HtrgGAT)} processes both spectral and temporal feature nodes. It projects each type of node, generates attention maps, and updates a master node that represents the aggregated features. Sequential layers are used to refine these features further. 
    \item The {graph pooling layer} reduces the number of nodes by selecting the top-k nodes based on attention scores. This process uses sigmoid activation and linear projection to compute the scores, with separate pooling layers for spectral and temporal features. 
    \item  The residual blocks apply convolutional layers, batch normalization, and SELU activation, similar to ResNet blocks, within the encoder to process input features. 
    \item  {The attention mechanism} derives spectral and temporal features from the encoded features, incorporating convolutional layers and SELU activation.
\end{itemize}

\subsubsection{Classifier}
The classifier outputs the final predictions by utilizing the refined features extracted from the backend model, subsequently performing the classification task. In this work, the input comprises a concatenation of maximum and average temporal features, maximum and average spectral features, and master node features from the ASSIST backend. To enhance generalization, dropout is applied to this concatenated feature vector. The output is generated through a linear layer, which produces logits, representing the raw scores.

\subsection{Model Ensembling }

Model ensembling is a strategy where multiple models are combined to improve the overall performance and robustness of predictions. The rationale behind this approach is that different models may capture various aspects of the data, and combining them can result in better generalization on unseen data. This method is widely adopted in many works in the anti-spoofing task~\cite{chettri19_interspeech, yang19b_interspeech}. In this work, we ensemble the individual models by averaging their output scores.

\section{Experimental Setup}
\label{sec:experiment}
\subsection{Data Set}

We utilized the official training and development datasets provided for the CtrSVDD track, available at Zenodo\footnote{\url{https://zenodo.org/records/10467648}}. Additionally, we incorporated other public datasets including JVS~\cite{jvs-music}, Kiritan~\cite{Kiritan}, Ofutan-P\footnote{\url{https://sites.google.com/view/oftn-utagoedb}}, and Oniku\footnote{\url{https://onikuru.info/db-download/}} following the guidelines and scripts provided by the challenge organizers~\cite{CtrSVDD_interspeech}. The combined dataset included a diverse range of singing voice recordings, both authentic and deepfake, segmented and processed\footnote{\url{https://github.com/SVDDChallenge/CtrSVDD_Utils}} to ensure consistency in training and evaluation. The details of the dataset partitions, along with the evaluation set statistics from~\cite{CtrSVDD_interspeech}, are provided in Table~\ref{tab:dataset}.
\begin{table}[h]
    \vspace{-0.15 in}

    \centering
    \caption{Dataset statistics.}
    \vspace{0.03 in}
    
    \label{tab:dataset}
    \begin{tabular}{llrr}
        \toprule
        \multirow{2}{*}{Partition} & \multirow{2}{*}{Speakers} & \multicolumn{2}{c}{ Utterances} \\
        \cmidrule(r){3-4}
         &  & Bonafide & Spoofed \\
        \midrule
        Train & 59 & 12,169 & 72,235 \\
        Dev & 55 & 6,547 & 37,078 \\
        Eval & 48 & 13,596 & 79,173 \\
        \bottomrule
    \end{tabular}
    \vspace{-0.15 in}
    
\end{table}

\subsection{Training Strategy}
We use the equal error rate (EER) as the evaluation metric. To ensure reproducibility, we consistently apply a fixed random seed of 42 across all systems. Our training process employs the AdamW optimizer with a batch size of 48, an initial learning rate of $1 \times 10^{-6}$, and a weight decay of $1 \times 10^{-4}$. The learning rate is scheduled using cosine annealing with a cycle to a minimum of $1 \times 10^{-9}$. For the loss function, we utilize a binary focal loss, a generalized form of binary cross-entropy loss, with a focusing parameter ($\gamma$) of 2 and a positive example weight ($\alpha$) of 0.25.  To standardize input length, each sample is randomly cropped or padded to 4 seconds during the training. Our model is trained for 30 epochs, and the model checkpoint with the lowest EER on the validation set is selected for evaluation. 
All experiments are performed on a single NVIDIA A100 GPU.

For certain experiments marked in Table~\ref{tab:model_performance}, we employ the Rawboost data augmentation strategy as introduced in Section~\ref{subsubsec:rawboost_aug}. 
The RawBoost augmentation is sourced from the official implementation\footnote{\url{https://github.com/TakHemlata/SSL_Anti-spoofing}} and  follows the default settings~\cite{hemlata_wav2vec2}. Our utilization of wav2vec2 also references this implementation. The wav2vec2~\cite{wav2vec2} model used in this work is the cross-lingual speech representations (XLSR) model\footnote{\url{https://github.com/facebookresearch/fairseq/tree/main/examples/wav2vec/xlsr}}. The implementation of WavLM is derived from S3PRL\footnote{\url{https://github.com/s3prl/s3prl}}.



\begin{table*}[t]
\centering
\caption{Performance in EER (\%) on the evaluation set of CtrSVDD for individual models. $\dag$ indicates re-implementation. All models use the AASIST backend. The table is best visualized in color mode, with darker red indicating higher EER and darker green indicating lower EER. For EER, smaller values indicate better performance. M9 and M10 are the best and second-best models from repeated experiments with different random seeds under the same settings. `(1)', `(2)', and `(3)' indicate the LnL Convolutive, ISD and SSI noise introduced in Section~\ref{subsubsec:rawboost_aug}.}
\vspace{0.06 in}
\resizebox{1.01\textwidth}{!}{%
\begin{tabular}{lcccrrrrrrrrrr}
\hline
 &  &  &  & \multicolumn{2}{c}{\textbf{EER of Datasets}} & \multicolumn{6}{c}{\textbf{EER of Different Attack Types}} & \multicolumn{2}{c}{\textbf{Pooled EER\footnotemark[8]}} \\ \cmidrule(r){5-6} \cmidrule(r){7-12} \cmidrule(r){13-14}
\multirow{-2}{*}{\textbf{Index}} & \multirow{-2}{*}{\textbf{Frontend}} & \multirow{-2}{*}{\textbf{\begin{tabular}[c]{@{}c@{}}Layer \\ Aggregation\end{tabular}}} & \multirow{-2}{*}{\textbf{Augmentation}} & \multicolumn{1}{c}{\textbf{\small m4singer}} & \multicolumn{1}{c}{\textbf{\small kising}} & \multicolumn{1}{c}{\textbf{\small A09}} & \multicolumn{1}{c}{\textbf{\small A10}} & \multicolumn{1}{c}{\textbf{\small A11}} & \multicolumn{1}{c}{\textbf{\small A12}} & \multicolumn{1}{c}{\textbf{\small A13}} & \multicolumn{1}{c}{\textbf{\small A14}} & \multicolumn{1}{l}{\footnotesize \textbf{A09-A14}} & \footnotesize \textbf{A09-A13}\\
\hline

B01~\cite{SVDD_challenge} & LFCCs & - & - & - & - & - & - & - & - & - & - & - & 11.37 \\
B02~\cite{SVDD_challenge} & Raw waveform & - & - & - & - & - & - & - & - & - & - & - & 10.39 \\
B01~\cite{CtrSVDD_interspeech} & LFCCs & - & - & - & - & \cellcolor[HTML]{FEF6F5}5.35 & \cellcolor[HTML]{E1F2EA}2.92 & \cellcolor[HTML]{FDF3F2}5.84 & \cellcolor[HTML]{E67C73}29.47 & \cellcolor[HTML]{FFFEFE}3.65 & \cellcolor[HTML]{EC9891}24.00 & 16.15 & - \\
B02~\cite{CtrSVDD_interspeech}  & Raw waveform & - & - & - & - & \cellcolor[HTML]{FCEFEE}6.72 & \cellcolor[HTML]{66C194}0.96 & \cellcolor[HTML]{FFFFFE}3.59 & \cellcolor[HTML]{E98A82}26.83 & \cellcolor[HTML]{65C094}{0.95} & \cellcolor[HTML]{F1B1AC}19.03 & 13.75 & - \\
B02$^{\dag}$ & Raw waveform & - & - & \cellcolor[HTML]{F8DAD8}10.77 & \cellcolor[HTML]{F8DBD8}10.73 & \cellcolor[HTML]{FDF2F1}6.14 & \cellcolor[HTML]{69C296}1.01 & \cellcolor[HTML]{FFFEFE}3.76 & \cellcolor[HTML]{EB968F}24.43 & \cellcolor[HTML]{74C69E}1.18 & \cellcolor[HTML]{F1B3AE}18.55 & 12.75 & 9.45 \\ \hline
M1 & wav2vec2 & - & - & \cellcolor[HTML]{FDF5F4}5.55 & \cellcolor[HTML]{F5CAC7}13.97 & \cellcolor[HTML]{B4E0CA}2.21 & \cellcolor[HTML]{9DD7BA}1.84 & \cellcolor[HTML]{FEF7F7}5.02 & \cellcolor[HTML]{FAE3E1}9.11 & \cellcolor[HTML]{CEEBDD}2.62 & \cellcolor[HTML]{F0B1AB}19.07 & 9.87 & 4.80 \\
M2 & wav2vec2 & - & series: (1)+(2) & \cellcolor[HTML]{FCEEED}6.83 & \cellcolor[HTML]{F9E0DE}9.71 & \cellcolor[HTML]{B1DFC8}2.16 & \cellcolor[HTML]{A9DCC3}2.03 & \cellcolor[HTML]{FAE5E3}8.71 & \cellcolor[HTML]{FCEEEC}6.95 & \cellcolor[HTML]{BCE4D0}2.34 & \cellcolor[HTML]{F6CCC9}13.57 & 7.94 & 5.99 \\
M3 & wav2vec2 & - & parallel: (1)+(2) & \cellcolor[HTML]{FFFDFD}3.94 & \cellcolor[HTML]{F9DEDC}10.00 & \cellcolor[HTML]{8DD1AF}1.59 & \cellcolor[HTML]{73C69D}1.17 & \cellcolor[HTML]{F1F9F5}3.19 & \cellcolor[HTML]{FCECEA}7.37 & \cellcolor[HTML]{9BD6B9}1.81 & \cellcolor[HTML]{F6CCC8}13.70 & 6.88 & 3.55 \\ \hline
M4 & WavLM & Weighted Sum & series: (1)+(2) & \cellcolor[HTML]{FEF9F9}4.68 & \cellcolor[HTML]{FAE4E2}8.81 & \cellcolor[HTML]{B4E0CA}2.21 & \cellcolor[HTML]{85CDAA}1.46 & \cellcolor[HTML]{FDF4F4}5.62 & \cellcolor[HTML]{FDF4F3}5.77 & \cellcolor[HTML]{91D2B2}1.66 & \cellcolor[HTML]{F6CFCC}12.98 & 6.66 & 4.10 \\
M5 & WavLM & Weighted Sum & parallel: (1)+(2) & \cellcolor[HTML]{FFFFFF}3.40 & \cellcolor[HTML]{FAE4E2}8.85 & \cellcolor[HTML]{7ECAA5}1.35 & \cellcolor[HTML]{67C195}0.98 & \cellcolor[HTML]{FFFEFE}3.70 & \cellcolor[HTML]{FDF4F3}5.78 & \cellcolor[HTML]{6CC399}1.07 & \cellcolor[HTML]{F7D2CF}12.52 & {5.91} & 3.16 \\ \cline{3-14} 
M6 & WavLM & AttM~\cite{attentive_merge} & series: (1)+(2) & \cellcolor[HTML]{FEF9F8}4.72 & \cellcolor[HTML]{F8D7D4}11.47 & \cellcolor[HTML]{93D3B3}1.68 & \cellcolor[HTML]{7BC9A3}1.29 & \cellcolor[HTML]{FDF0EF}6.44 & \cellcolor[HTML]{FDF0EF}6.44 & \cellcolor[HTML]{88CFAC}1.51 & \cellcolor[HTML]{F5C7C3}14.67 & 7.63 & 4.26 \\
M7 & WavLM & AttM~\cite{attentive_merge} & parallel: (1)+(2) & \cellcolor[HTML]{FFFFFF}3.48 & \cellcolor[HTML]{F8DBD8}10.73 & \cellcolor[HTML]{74C79E}\textbf{1.19} & \cellcolor[HTML]{57BB8A}\textbf{0.72} & \cellcolor[HTML]{FFFDFD}3.81 & \cellcolor[HTML]{FDF2F1}6.02 & \cellcolor[HTML]{60BE90}\textbf{0.87} & \cellcolor[HTML]{F6CCC8}13.70 & 6.51 & 3.22 \\ \cline{3-14} 
M8 & WavLM & Proposed SEA & series: (1)+(2) & \cellcolor[HTML]{FFFDFD}3.81 & \cellcolor[HTML]{FBE6E4}8.53 & \cellcolor[HTML]{7CCAA4}{1.32} & \cellcolor[HTML]{64C093}{0.93} & \cellcolor[HTML]{FFFEFE}3.72 & \cellcolor[HTML]{FDF3F2}5.95 & \cellcolor[HTML]{71C59C}1.15 & \cellcolor[HTML]{F6D0CD}12.83 & 6.16 & 3.32 \\
M9 & WavLM & Proposed SEA & parallel: (1)+(2) & \cellcolor[HTML]{DCF0E6}\textbf{2.84} & \cellcolor[HTML]{FBE7E5}8.36 & \cellcolor[HTML]{8FD1B1}1.62 & \cellcolor[HTML]{76C7A0}1.23 & \cellcolor[HTML]{BDE4D1}\textbf{2.35} & \cellcolor[HTML]{FEF6F6}{5.24} & \cellcolor[HTML]{7CCAA4}1.32 & \cellcolor[HTML]{F7D2CF}{12.46} & \textbf{5.66} & \textbf{2.70} \\
M10 & WavLM & Proposed SEA & parallel: (1)+(2) & \cellcolor[HTML]{F6FBF9}{3.26} & \cellcolor[HTML]{FAE1DE}9.54 & \cellcolor[HTML]{89CFAD}1.52 & \cellcolor[HTML]{6CC398}1.06 & \cellcolor[HTML]{D0ECDE}{2.66} & \cellcolor[HTML]{FDF3F2}5.98 & \cellcolor[HTML]{72C69D}1.16 & \cellcolor[HTML]{F6D0CC}12.91 & 5.94 & {3.02} \\
M11 & WavLM & Proposed SEA & series: (1)+(3) & \cellcolor[HTML]{FCF0EE}6.57 & \cellcolor[HTML]{FEF7F7}{5.03} & \cellcolor[HTML]{C4E7D6}2.47 & \cellcolor[HTML]{99D6B8}1.79 & \cellcolor[HTML]{FAE1DF}9.53 & \cellcolor[HTML]{FEF7F6}\textbf{5.10} & \cellcolor[HTML]{A5DAC0}1.97 & \cellcolor[HTML]{F7D3CF}\textbf{12.35} & 7.36 & 5.77 \\
M12 & WavLM & Proposed SEA & series: (2)+(3) & \cellcolor[HTML]{FCECEB}7.24 & \cellcolor[HTML]{FEF7F7}\textbf{5.00} & \cellcolor[HTML]{D4EDE1}2.71 & \cellcolor[HTML]{B7E2CD}2.26 & \cellcolor[HTML]{FAE5E3}8.70 & \cellcolor[HTML]{FCEFEE}6.66 & \cellcolor[HTML]{C4E7D5}2.46 & \cellcolor[HTML]{F6CCC9}13.56 & 7.76 & 6.08
\\
\bottomrule
\end{tabular}
}
\vspace{-0.15 in}
\label{tab:model_performance}
\end{table*}

\begin{table*}[t]

\centering
\caption{Performance in EER (\%) on the evaluation set of CtrSVDD for ensemble systems.}
\vspace{0.06 in}
\resizebox{1.01\textwidth}{!}{%

\begin{tabular}{clccccccccccc}
\hline
\multirow{2}{*}{\textbf{Index}} & \multicolumn{1}{c}{\multirow{2}{*}{\textbf{Ensembling Details}}} & \multirow{2}{*}{\textbf{\begin{tabular}[c]{@{}c@{}}Ensemble \\ Adjustments\end{tabular}}} & \multicolumn{2}{c}{\textbf{EER of Datasets}} & \multicolumn{6}{c}{\textbf{EER of Different Attackers}} & \multicolumn{2}{c}{\textbf{Pooled EER\footnotemark[8]}} \\ \cmidrule(r){4-5} \cmidrule(r){6-11} \cmidrule(r){12-13}
 & \multicolumn{1}{c}{} &  & \multicolumn{1}{c}{\textbf{m4singer}} & \multicolumn{1}{c}{\textbf{kising}} & \multicolumn{1}{c}{\textbf{A09}} & \multicolumn{1}{c}{\textbf{A10}} & \multicolumn{1}{c}{\textbf{A11}} & \multicolumn{1}{c}{\textbf{A12}} & \multicolumn{1}{c}{\textbf{A13}} & \multicolumn{1}{c}{\textbf{A14}} & \multicolumn{1}{c}{\textbf{\small A09-A14}} & \multicolumn{1}{c}{\textbf{\small A09-A13}} \\ \cline{1-13}
E1 & M5 + M7 + M8 + M9 + M10 & - & 2.71 & 8.40 & 1.03 & 0.74 & 2.56 & 4.77 & 0.88 & 12.33 & 5.39 & 2.50 \\
E2 & M3 + M5 + M7 + M8 + M9 + M10 & +M3 & 2.41 & 7.19 & 0.82 & 0.56 & 2.17 & 4.24 & 0.69 & 12.00 & 5.01 & 2.21 \\
E3 & M3 + M5 + M7 + M9 + M10 & -M8 & 2.30 & 7.21 & 0.79 & 0.55 & 2.00 & 4.17 & 0.70 & 11.94 & 4.96 & 2.13 \\
E4 & M2 + M3 + M5 + M7 + M9 + M10 & +M2 & 2.09 & 6.47 & 0.68 & 0.48 & 1.96 & 3.83 & 0.63 & \textbf{11.80} & 4.78 & 1.95 \\
E5 & M2 + M3 + M7 + M9 + M10 & -M5 & \textbf{1.93} & \textbf{6.02} & \textbf{0.58} & \textbf{0.44} & \textbf{1.67} & \textbf{3.82} & \textbf{0.56} & 11.84 & \textbf{4.76} & \textbf{1.79} \\
\hline
\end{tabular}
}
\label{tab:ensembled_models}
\vspace{-0.05 in}
\end{table*}

\section{Results}
\label{sec:Results}
\subsection{Baselines}

The organizers of the CtrSVDD Challenge 2024 provide two baseline systems, referred to as B01 and B02 in Table~\ref{tab:model_performance}~\cite{SVDD_challenge, CtrSVDD_interspeech}.  B01, based on linear frequency cepstral coefficients (LFCCs), achieved a pooled EER of 11.37\%, while B02, based on raw waveform, achieved a pooled EER of 10.39\%. We re-implement B02 and obtain an improved performance of 9.45\%, slightly better than the official implementation.

\subsection{Frontend}

{As indicated  in Table~\ref{tab:model_performance}, when comparing wav2vec2-based models to WavLM-based models with the same type of augmentation (M2 vs. M4 for RawBoost `series: (1)+(2)', and M3 vs. M5 for  `parallel: (1)+(2)'), we observe that the WavLM-based models consistently perform better. Therefore, in this work, we focus more on experimenting with WavLM-based models.}

\subsection{Data Augmentation}

By comparing the wav2vec2-based models trained with and without `parallel: (1)+(2)' RawBoost augmentation~\cite{rawboost} (M1 vs. M3), we observe a significant improvement in performance when the augmentation is applied. Further analysis based on various models and layer aggregation techniques reveals that the `parallel: (1)+(2)' configuration consistently provides better results compared to the `series: (1)+(2)' configuration (M2 vs. M3, M4 vs. M5, M6 vs. M7, M8 vs. M9), with an average relative performance improvement of 26.7\%.
On the other hand, our experiments show that using type (3) of RawBoost (SSI additive noise)~\cite{rawboost} does not yield more benefits (M11 and M12). Overall, RawBoost generally enhances system performance on the CtrSVDD dataset. Notably, benefiting from `parallel: (1)+(2)', the WavLM-based model with our proposed SEA (M9) achieves the best individual performance on the evaluation set, as shown in Table~\ref{tab:model_performance}.

\footnotetext[8]{We report the overall system performance according to the settings in the SVDD Challenge 2024~\cite{SVDD_challenge}, which calculates the pooled Equal Error Rate (EER) for attack types A09 to A13, excluding A14. Additionally, for the benefit of interested readers, we also include the pooled EER results for all attack types (A09 to A14).}



\subsection{Layer Aggregation Strategies}
As shown in Table~\ref{tab:model_performance},when comparing different layer aggregation methods, we observe that the AttM strategy performs similarly to the weighted sum method in terms of pooled EER. Additionally, the AttM model (M7) achieves the best performance in the most sub-trials.
In this work, we simply utilize all WavLM layers, while the strength of AttM method lies in using fewer encoder layers. This not only lowers inference costs but also boosts performance~\cite{attentive_merge}. This aspect is valuable for exploring in the SVDD task. 

Given that the weighted sum method lacks a cross-layer attention mechanism, which may limit the representation features extracted by the speech foundation model in complex musical scenarios, and that AttM's higher number of training parameters could lead to overfitting on small datasets, we propose the SEA method. 
Our proposed SEA aggregation method, based on the WavLM model, consistently outperforms both the Weighted Sum and AttM across different RawBoost augmentation scenarios, achieving an average relative reduction in EER by 16.7\% and 19.1\%, respectively.
With this proposed SEA, we achieve the best individual model performance of 2.70\%, validating its superior performance and suitability for the task of singing voice deepfake detection.

\subsection{Model Ensembling} 
\begin{figure}[h]
\centerline{\includegraphics[scale=0.68]{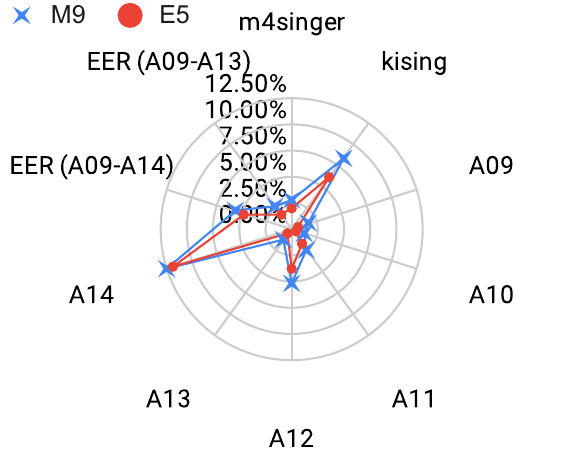}}
\caption{The radar chart comparing the performance of our best individual model (M9) and the best ensemble system (E5) in terms of EER on sub-trials of the CtrSVDD evaluation set.}
\label{fig_radar}
\end{figure}
\vspace{0.2 in}

We explore ensembling models to enhance robustness and performance. The ensembled models and their corresponding evaluation EER are shown in Table \ref{tab:ensembled_models}.  Specifically,
{we explore the model ensembling strategy by initially ensembling the top 5 individual models based on their performance on A09-A14 pooled EER. The E1 system, composed of M5, M7, M8, M9, and M10, achieves a 2.50\% EER, outperforming all individual systems. Further investigation includes incorporating a wav2vec2-based model to enhance system diversity and robustness improvement. Consequently, we include the best wav2vec2 system, M3, and remove the weakest individual model, M8, from E1, resulting in E3, which performs at 2.13\%. During post-evaluation, we further improve the ensemble performance by adding M2 and removing M5, achieving the best performance of 1.79\%.}

We note that although the pooled EER of the M2 model is not as good as other models in Table 2, it significantly contributes to ensemble performance. Since the evaluation labels have not yet been released, further analysis is not possible in this study. However, future investigations will help in understanding this improvement.

In Fig.~\ref{fig_radar}, we provide a detailed comparison of the best individual model, the WavLM-based model with our proposed SEA (M9), and the best ensemble system (E5). The radar chart clearly illustrates that E5 consistently outperforms M9 in every sub-trial.
This demonstrates the superiority and robustness of ensemble systems by combining the strengths of multiple models, reducing the impact of individual model errors, and increasing overall prediction accuracy.

\section{Conclusion}
\label{sec:conclusion}

In this work, we present ensembled systems utilizing speech foundation models, demonstrating  significant promise in the task of singing voice deepfake detection (SVDD). 
Our novel layer aggregation strategy, SE Aggregation (SEA), enables the WavLM-based model to achieve the best performance with a 2.70\% EER on the CtrSVDD evaluation set, outperforming all individual models. By implementing data augmentation techniques, such as RawBoost, our ensembled system further achieves a remarkable 1.79\% pooled EER on the CtrSVDD evaluation set.
Further analysis validates that model ensembling effectively combines the strengths of different models, enhancing both robustness and accuracy. These findings contribute to advancing the field of audio anti-spoofing, particularly in SVDD. Future work can explore further optimization of layer aggregation techniques and broader applications to improve detection systems.

\section{Acknowledgements}
This work is supported
by the National Research Foundation, Prime Minister’s Office, Singapore, and the Ministry of Digital Development and Information, under its Online Trust and Safety (OTS) Research Programme (MCI-OTS-001). Any opinions, findings and conclusions or recommendations expressed in this material are those of the author(s) and do not reflect the views of National Research Foundation, Prime Minister’s Office,  Singapore, or the Ministry of Digital Development and Information.





\bibliographystyle{IEEEbib_limit6}
\bibliography{refs}

\begin{thebibliography}{10}

\bibitem{7858696}
Zhizheng Wu, Junichi Yamagishi, Tomi Kinnunen, Cemal Hanilçi, Mohammed Sahidullah et~al.,
\newblock ``Asvspoof: The automatic speaker verification spoofing and countermeasures challenge,''
\newblock {\em IEEE Journal of Selected Topics in Signal Processing}, vol. 11, no. 4, pp. 588--604, 2017.

\bibitem{ASVspoof2019}
Massimiliano Todisco, Xin Wang, Ville Vestman, Md~Sahidullah, H{\'e}ctor Delgado et~al.,
\newblock ``{{ASV}spoof 2019: Future Horizons in Spoofed and Fake Audio Detection},''
\newblock in {\em Proc. Interspeech}, 2019, pp. 1008--1012.

\bibitem{delgado2024asvspoof}
H{\'e}ctor Delgado, Nicholas Evans, Jee-weon Jung, Tomi Kinnunen, Ivan Kukanov et~al.,
\newblock ``Asvspoof 5 evaluation plan,''
\newblock 2024.

\bibitem{one_class}
You Zhang, Fei Jiang and Zhiyao Duan,
\newblock ``One-class learning towards synthetic voice spoofing detection,''
\newblock {\em IEEE Signal Processing Letters}, vol. 28, pp. 937--941, 2021.

\bibitem{AASIST}
Jee-weon Jung, Hee-Soo Heo, Hemlata Tak, Hye-jin Shim, Joon~Son Chung et~al.,
\newblock ``Aasist: Audio anti-spoofing using integrated spectro-temporal graph attention networks,''
\newblock in {\em Proc. ICASSP}, 2022, pp. 6367--6371.

\bibitem{wu2024codecfake}
Haibin Wu, Yuan Tseng and Hung-yi Lee,
\newblock ``Codecfake: Enhancing anti-spoofing models against deepfake audios from codec-based speech synthesis systems,''
\newblock {\em arXiv preprint arXiv:2406.07237}, 2024.

\bibitem{SingFake}
Yongyi Zang, You Zhang, Mojtaba Heydari and Zhiyao Duan,
\newblock ``Singfake: Singing voice deepfake detection,''
\newblock in {\em Proc. ICASSP}, 2024, pp. 12156--12160.

\bibitem{CtrSVDD_interspeech}
Yongyi Zang, Jiatong Shi, You Zhang, Ryuichi Yamamoto, Jionghao Han et~al.,
\newblock ``Ctrsvdd: A benchmark dataset and baseline analysis for controlled singing voice deepfake detection,''
\newblock {\em arXiv preprint arXiv:2406.02438}, 2024.

\bibitem{10446271}
Yuankun Xie, Jingjing Zhou, Xiaolin Lu, Zhenghao Jiang, Yuxin Yang et~al.,
\newblock ``Fsd: An initial chinese dataset for fake song detection,''
\newblock in {\em Proc. ICASSP}, 2024, pp. 4605--4609.

\bibitem{SVDD_challenge}
You Zhang, Yongyi Zang, Jiatong Shi, Ryuichi Yamamoto, Jionghao Han et~al.,
\newblock ``Svdd challenge 2024: A singing voice deepfake detection challenge evaluation plan,''
\newblock {\em arXiv preprint arXiv:2405.05244}, 2024.

\bibitem{chen2024singing}
Xuanjun Chen, Haibin Wu, Jyh-Shing~Roger Jang and Hung yi~Lee,
\newblock ``Singing voice graph modeling for singfake detection,'' 2024.

\bibitem{zhang2024svdd2024inauguralsinging}
You Zhang, Yongyi Zang, Jiatong Shi, Ryuichi Yamamoto, Tomoki Toda and Zhiyao Duan,
\newblock ``Svdd 2024: The inaugural singing voice deepfake detection challenge,''
\newblock {\em arXiv preprint arXiv:2408.16132}, 2024.

\bibitem{lin2024sawavlms}
Jingru Lin, Meng Ge, Junyi Ao, Liqun Deng and Haizhou Li,
\newblock ``Sa-wavlm: Speaker-aware self-supervised pre-training for mixture speech,''
\newblock {\em arXiv preprint arXiv:2407.02826}, 2024.

\bibitem{10446072}
Yidi Jiang, Zhengyang Chen, Ruijie Tao, Liqun Deng, Yanmin Qian and Haizhou Li,
\newblock ``Prompt-driven target speech diarization,''
\newblock in {\em Proc. ICASSP}, 2024, pp. 11086--11090.

\bibitem{jiang2024target}
Yidi Jiang, Ruijie Tao, Zhengyang Chen, Yanmin Qian and Haizhou Li,
\newblock ``Target speech diarization with multimodal prompts,''
\newblock {\em arXiv preprint arXiv:2406.07198}, 2024.

\bibitem{wavlm}
Sanyuan Chen, Chengyi Wang, Zhengyang Chen, Yu~Wu, Shujie Liu et~al.,
\newblock ``Wavlm: Large-scale self-supervised pre-training for full stack speech processing,''
\newblock {\em IEEE Journal of Selected Topics in Signal Processing}, vol. 16, no. 6, pp. 1505--1518, 2022.

\bibitem{wav2vec2}
Alexei Baevski, Yuhao Zhou, Abdelrahman Mohamed and Michael Auli,
\newblock ``wav2vec 2.0: A framework for self-supervised learning of speech representations,''
\newblock in {\em Proc. NeurIPS}, 2020, vol.~33, pp. 12449--12460.

\bibitem{10003971}
Lin Zhang, Xin Wang, Erica Cooper, Nicholas Evans and Junichi Yamagishi,
\newblock ``{The PartialSpoof Database and Countermeasures for the Detection of Short Fake Speech Segments Embedded in an Utterance},''
\newblock {\em IEEE/ACM Transactions on Audio, Speech, and Language Processing}, vol. 31, pp. 813--825, 2023.

\bibitem{liu2024neural}
Tianchi Liu, Lin Zhang, Rohan~Kumar Das, Yi~Ma, Ruijie Tao and Haizhou Li,
\newblock ``How do neural spoofing countermeasures detect partially spoofed audio?,''
\newblock {\em arXiv preprint arXiv:2406.02483}, 2024.

\bibitem{10446331}
Xin Wang and Junichi Yamagishi,
\newblock ``Can large-scale vocoded spoofed data improve speech spoofing countermeasure with a self-supervised front end?,''
\newblock in {\em Proc. ICASSP}, 2024, pp. 10311--10315.

\bibitem{9747768}
Juan~M. Martín-Doñas and Aitor Álvarez,
\newblock ``The vicomtech audio deepfake detection system based on wav2vec2 for the 2022 add challenge,''
\newblock in {\em Proc. ICASSP}, 2022, pp. 9241--9245.

\bibitem{10448049}
Yuxiang Zhang, Jingze Lu, Zengqiang Shang, Wenchao Wang and Pengyuan Zhang,
\newblock ``Improving short utterance anti-spoofing with aasist2,''
\newblock in {\em Proc. ICASSP}, 2024, pp. 11636--11640.

\bibitem{10448016}
Wanying Ge, Xin Wang, Junichi Yamagishi, Massimiliano Todisco and Nicholas Evans,
\newblock ``Spoofing attack augmentation: Can differently-trained attack models improve generalisation?,''
\newblock in {\em Proc. ICASSP}, 2024, pp. 12531--12535.

\bibitem{attentive_merge}
Zihan Pan, Tianchi Liu, Hardik~B Sailor and Qiongqiong Wang,
\newblock ``Attentive merging of hidden embeddings from pre-trained speech model for anti-spoofing detection,''
\newblock {\em arXiv preprint arXiv:2406.10283}, 2024.

\bibitem{SE}
Jie Hu, Li~Shen and Gang Sun,
\newblock ``Squeeze-and-excitation networks,''
\newblock in {\em Proc. CVPR}, June 2018.

\bibitem{rawboost}
Hemlata Tak, Madhu Kamble, Jose Patino, Massimiliano Todisco and Nicholas Evans,
\newblock ``Rawboost: A raw data boosting and augmentation method applied to automatic speaker verification anti-spoofing,''
\newblock in {\em Proc. ICASSP}, 2022, pp. 6382--6386.

\bibitem{jung20c_interspeech}
Jee weon Jung, Seung bin Kim, Hye jin Shim, Ju~ho~Kim and Ha-Jin Yu,
\newblock ``{Improved RawNet with Feature Map Scaling for Text-Independent Speaker Verification Using Raw Waveforms},''
\newblock in {\em Proc. Interspeech}, 2020, pp. 1496--1500.

\bibitem{truong2024temporal}
Duc-Tuan Truong, Ruijie Tao, Tuan Nguyen, Hieu-Thi Luong, Kong~Aik Lee and Eng~Siong Chng,
\newblock ``Temporal-channel modeling in multi-head self-attention for synthetic speech detection,''
\newblock {\em arXiv preprint arXiv:2406.17376}, 2024.

\bibitem{yang21c_interspeech}
Shu wen Yang, Po-Han Chi, Yung-Sung Chuang, Cheng-I~Jeff Lai, Kushal Lakhotia et~al.,
\newblock ``{SUPERB: Speech Processing Universal PERformance Benchmark},''
\newblock in {\em Proc. Interspeech}, 2021, pp. 1194--1198.

\bibitem{huo23b_interspeech}
Zhouyuan Huo, Khe~Chai Sim, Dongseong Hwang, Tsendsuren Munkhdalai, Tara Sainath and Pedro~M. Mengibar,
\newblock ``{Re-investigating the Efficient Transfer Learning of Speech Foundation Model using Feature Fusion Methods},''
\newblock in {\em Proc. Interspeech}, 2023, pp. 556--560.

\bibitem{MFA_TDNN}
Tianchi Liu, Rohan~Kumar Das, Kong~Aik Lee and Haizhou Li,
\newblock ``{MFA}: {TDNN} with multi-scale frequency-channel attention for text-independent speaker verification with short utterances,''
\newblock in {\em Proc. ICASSP}, 2022, pp. 7517--7521.

\bibitem{lai19b_interspeech}
Cheng-I Lai, Nanxin Chen, Jesús Villalba and Najim Dehak,
\newblock ``{ASSERT: Anti-Spoofing with Squeeze-Excitation and Residual Networks},''
\newblock in {\em Proc. Interspeech}, 2019, pp. 1013--1017.

\bibitem{9003763}
Songxiang Liu, Haibin Wu, Hung-Yi Lee and Helen Meng,
\newblock ``Adversarial attacks on spoofing countermeasures of automatic speaker verification,''
\newblock in {\em Proc. ASRU}, 2019, pp. 312--319.

\bibitem{Desplanques20}
Brecht Desplanques, Jenthe Thienpondt and Kris Demuynck,
\newblock ``{ECAPA-TDNN}: Emphasized channel attention, propagation and aggregation in {TDNN} based speaker verification,''
\newblock in {\em Proc. Interspeech}, 2020, pp. 3830--3834.

\bibitem{RecXi}
Tianchi Liu, Kong~Aik Lee, Qiongqiong Wang and Haizhou Li,
\newblock ``Disentangling voice and content with self-supervision for speaker recognition,''
\newblock in {\em Proc. NeurIPS}, 2023, vol.~36, pp. 50221--50236.

\bibitem{liu2023golden}
Tianchi Liu, Kong~Aik Lee, Qiongqiong Wang and Haizhou Li,
\newblock ``{Golden Ge}mini is all you need: Finding the sweet spots for speaker verification,''
\newblock {\em IEEE/ACM Transactions on Audio, Speech, and Language Processing}, vol. 32, pp. 2324--2337, 2024.

\bibitem{chettri19_interspeech}
Bhusan Chettri, Daniel Stoller, Veronica Morfi, Marco A.~Martínez Ramírez, Emmanouil Benetos and Bob~L. Sturm,
\newblock ``{Ensemble Models for Spoofing Detection in Automatic Speaker Verification},''
\newblock in {\em Proc. Interspeech}, 2019, pp. 1018--1022.

\bibitem{yang19b_interspeech}
Yexin Yang, Hongji Wang, Heinrich Dinkel, Zhengyang Chen, Shuai Wang et~al.,
\newblock ``{The SJTU Robust Anti-Spoofing System for the ASVspoof 2019 Challenge},''
\newblock in {\em Proc. Interspeech}, 2019, pp. 1038--1042.

\bibitem{jvs-music}
Hiroki Tamaru, Shinnosuke Takamichi, Naoko Tanji and Hiroshi Saruwatari,
\newblock ``Jvs-music: Japanese multispeaker singing-voice corpus,''
\newblock {\em arXiv preprint arXiv:2001.07044}, 2020.

\bibitem{Kiritan}
Itsuki Ogawa and Masanori Morise,
\newblock ``Tohoku kiritan singing database: A singing database for statistical parametric singing synthesis using japanese pop songs,''
\newblock {\em Acoustical Science and Technology}, vol. 42, no. 3, pp. 140--145, 2021.

\bibitem{hemlata_wav2vec2}
Hemlata Tak, Massimiliano Todisco, Xin Wang, Jee-weon Jung, Junichi Yamagishi and Nicholas Evans,
\newblock ``{Automatic Speaker Verification Spoofing and Deepfake Detection Using wav2vec 2.0 and Data Augmentation},''
\newblock in {\em Proc. Odyssey}, 2022, pp. 112--119.

\end{thebibliography}

\end{document}